\documentclass[twocolumn, 10pt, a4paper]{article}

\usepackage[utf8]{inputenc}
\usepackage[english]{babel}
\usepackage{amsmath, amssymb, amsfonts} 
\usepackage{geometry} 
\usepackage{hyperref} 
\usepackage{cite} 
\usepackage{authblk} 
\usepackage{graphicx} 
\usepackage{physics}
\usepackage{amsthm}

\theoremstyle{plain}
\newtheorem{theorem}{Theorem}[section]
\newtheorem{proposition}[theorem]{Proposition}
\newtheorem{corollary}[theorem]{Corollary}

\theoremstyle{definition}

\theoremstyle{remark}

\title{Probability Conservation, Liouville Measure, and the Symplectic Origin of Hamiltonian Dynamics}

\author[1]{Enmanuel Rodr\'iguez-Brea}
\author[1,2]{Melvin Arias}

\affil[1]{\textit{\small Instituto de F\'isica, Universidad Aut\'onoma de Santo Domingo, Av. Alma Mater, Santo Domingo 10105, Dominican Republic}}
\affil[2]{\textit{\small Laboratorio de Nanotecnolog\'ia, \'Area de Ciencias B\'asicas y Ambientales, Instituto Tecnol\'ogico de Santo Domingo, Av. Los Pr\'oceres, Santo Domingo 10602, Dominican Republic}}

\date{\today}

\begin{document}

\maketitle

\begin{abstract}
Liouville’s theorem---the preservation of phase-space volume---is often presented as a corollary of
Hamilton’s canonical equations. Here we adopt an ensemble-first viewpoint in which the starting point
is local probability conservation on phase space. For a probability density $\rho$ on a
$2N$-dimensional symplectic manifold $(\mathcal{M},\omega)$, probability transport is expressed intrinsically
with respect to the Liouville volume form $\Omega=\omega^N/N!$ through a continuity equation defined
by the $\Omega$-divergence. For Hamiltonian evolution, specified by $\iota_{X_H}\omega=\dd H$,
Cartan’s identity implies $\mathcal{L}_{X_H}\omega=0$ and hence $\mathcal{L}_{X_H}\Omega=0$, so the
Hamiltonian flow is incompressible in the Liouville sense and the continuity law reduces to
Liouville’s equation. In canonical coordinates this reproduces Hamilton’s equations. In particular,
the canonical Poisson-bracket relations $\{q^i,p_j\}=\delta^i_{\ j}$ provide the kinematic input that
fixes the evolution of observables and underlies the canonical form of the continuity equation. The
same organization clarifies the distinction between conservation of total probability and
preservation of fine-grained information measures (Gibbs--Shannon entropy), which holds specifically
for Liouville-measure-preserving dynamics.
\end{abstract}


\section{Introduction}

Hamiltonian mechanics is often introduced in a purely dynamical way: one postulates canonical
variables $(q^i,p_i)$, writes Hamilton's equations, and only afterward proves Liouville's theorem as
a corollary by computing the divergence of the associated phase-space velocity field
\cite{goldstein,arnold}. While mathematically correct, this ordering can obscure a structural point
that is central to both statistical mechanics and geometric formulations of classical physics;
Hamiltonian flows form a distinguished class of dynamics that naturally preserve this measure due to the symplectic structure.

In this paper we emphasize a complementary perspective that starts from ensembles rather than single
trajectories \cite{HenrikssonLJP2022}. We consider a probability density $\rho(\xi,t)$ on a $2N$-dimensional symplectic
manifold $(\mathcal{M},\omega)$ and impose local probability conservation through a continuity equation.
To avoid coordinate-dependent ambiguities, the continuity law is formulated intrinsically with
respect to the Liouville volume form $\Omega=\omega^N/N!$, using the divergence defined by
$\mathcal{L}_X\Omega=(\mathrm{div}_{\Omega}X)\Omega$. This makes clear which statements are purely
geometric and which depend on a particular choice of coordinates.

A second pedagogical motivation concerns the canonical form of Hamilton's equations. Students often
treat the momentum equation $\dot p_i=-\partial H/\partial q^i$ and its characteristic minus sign as
a convention that must be memorized. In the symplectic viewpoint, that sign is not arbitrary; it is
fixed by the antisymmetry of $\omega$ and the canonical pairing encoded in
$\omega=\sum_i \dd q^i\wedge \dd p_i$. Moreover, once $\omega$ and a Hamiltonian function $H$ are
specified, the Hamiltonian vector field $X_H$ is uniquely determined by $\iota_{X_H}\omega=\dd H$.
From this defining relation it follows that $\mathcal{L}_{X_H}\omega=0$ and thus
$\mathcal{L}_{X_H}\Omega=0$, yielding incompressibility in the Liouville sense and the Liouville
equation for $\rho$.

This organization also clarifies the meaning of ``information conservation'' in phase space.
The continuity equation guarantees conservation of total probability (normalization) for any smooth
flow under appropriate boundary conditions, but it does not by itself imply preservation of
fine-grained information measures. Fine-grained Gibbs--Shannon entropy is conserved specifically for
Liouville-measure-preserving dynamics (with Hamiltonian flows as the canonical case), whereas
dissipative systems may still satisfy a continuity equation while failing to preserve $\Omega$,
leading to phase-space contraction or expansion.

Beyond its conceptual value, the present route provides a natural bridge to statistical mechanics,
where Liouville's equation is the starting point for the derivation of equilibrium and nonequilibrium
ensembles \cite{pathria}, and it connects cleanly to quantum mechanics, where the commutator plays
the role of the Poisson bracket in the Heisenberg and von Neumann equations \cite{KoidePRA2021,sakurai,dirac}.
Our aim is therefore both foundational and pedagogical: to show how probability transport and
symplectic geometry jointly organize the standard Hamiltonian formalism, and to present the core
results in a compact form that is readily usable in advanced mechanics courses.


\section{Mathematical Framework}

We summarize the geometric and statistical ingredients used throughout the paper. Let $(\mathcal{M},\omega)$ be
a $2N$-dimensional symplectic manifold, with local coordinates $\xi^a$ ($a=1,\dots,2N$). Observables
are smooth functions $F\in C^\infty(\mathcal{M})$.

\subsection{Symplectic structure and Poisson bracket}

The symplectic form $\omega$ is a closed, nondegenerate $2$-form. In local coordinates,
\begin{equation}
\omega=\frac{1}{2}\,\omega_{ab}(\xi)\,\dd \xi^a\wedge \dd \xi^b,
\qquad
\dd\omega=0,
\qquad
\det(\omega_{ab})\neq 0.
\label{eq:omega_local}
\end{equation}
Nondegeneracy implies the existence of the inverse Poisson tensor $J^{ab}(\xi)$ defined by
\begin{equation}
\omega_{ac}J^{cb}=\delta_a^{\ b}.
\label{eq:inverse_relation}
\end{equation}
The associated Poisson bracket on $C^\infty(\mathcal{M})$ is
\begin{equation}
\{F,G\}:=J^{ab}\,\partial_a F\,\partial_b G,
\qquad \partial_a:=\pdv{}{\xi^a},
\label{eq:PB_general}
\end{equation}
which is bilinear, antisymmetric, and satisfies the Leibniz rule. Since $\dd\omega=0$, the bracket
obeys the Jacobi identity, making $C^\infty(\mathcal{M}),\{\cdot,\cdot\}$ a Poisson algebra.

In canonical coordinates $\xi^a=(q^1,\dots,q^N,p_1,\dots,p_N)$, the symplectic form reads
\begin{equation}
\omega=\sum_{i=1}^N \dd q^i\wedge \dd p_i,
\label{eq:omega_canonical}
\end{equation}
and the fundamental Poisson brackets become
\begin{equation}
\{q^i,q^j\}=0,
\qquad
\{p_i,p_j\}=0,
\qquad
\{q^i,p_j\}=\delta^i_{\ j}.
\label{eq:PB_canonical}
\end{equation}

For an operator-based perspective on the Poisson bracket, see Ref.~\cite{KoidePRA2021}.

\subsection{Hamiltonian vector field and evolution of observables}

Given a Hamiltonian function $H\in C^\infty(\mathcal{M})$, the Hamiltonian vector field $X_H$ is defined
uniquely by
\begin{equation}
\iota_{X_H}\omega=\dd H,
\label{eq:XH_def}
\end{equation}
where $\iota_{X_H}$ denotes interior contraction. The induced evolution of an observable $F(\xi,t)$ is
\begin{equation}
\frac{\dd F}{\dd t}=X_H[F]+\pdv{F}{t}=\{F,H\}+\pdv{F}{t}.
\label{eq:observable_evolution}
\end{equation}
When $q^i$ and $p_i$ have no explicit time dependence, \eqref{eq:observable_evolution} yields the
canonical Hamilton equations,
\begin{equation}
\dot q^i=\{q^i,H\}=\pdv{H}{p_i},
\qquad
\dot p_i=\{p_i,H\}=-\pdv{H}{q^i}.
\label{eq:Hamilton_eqs}
\end{equation}

\subsection{Liouville volume form and intrinsic continuity equation}

To describe an ensemble, we introduce a phase-space probability density $\rho(\xi,t)\ge 0$.
Probabilities are computed with respect to the Liouville volume form
\begin{equation}
\Omega:=\frac{\omega^N}{N!}.
\label{eq:Liouville_volume_form}
\end{equation}
Normalization is
\begin{equation}
\int_\mathcal{M} \rho(\xi,t)\,\Omega = 1.
\label{eq:normalization}
\end{equation}

For a general phase-space flow generated by a vector field $X$, not necessarily Hamiltonian,
local probability conservation is encoded by a continuity equation. The appropriate divergence is the one
defined by the volume form $\Omega$:
\begin{equation}
\mathcal{L}_X \Omega = \big(\mathrm{div}_{\Omega}X\big)\,\Omega,
\label{eq:divOmega_def}
\end{equation}
where $\mathcal{L}_X$ denotes the Lie derivative. The intrinsic continuity equation then reads
\begin{equation}
\pdv{\rho}{t}+\mathrm{div}_{\Omega}(\rho X)=0.
\label{eq:continuity_intrinsic}
\end{equation}

In canonical coordinates, where $\Omega=\dd^Nq\,\dd^Np$, Eq.~\eqref{eq:continuity_intrinsic} reduces to
\begin{equation}
\pdv{\rho}{t}+\sum_{i=1}^N\left[
\pdv{}{q^i}\!\big(\rho\,\dot q^i\big)+
\pdv{}{p_i}\!\big(\rho\,\dot p_i\big)\right]=0.
\label{eq:continuity_canonical}
\end{equation}
A sufficient condition for global probability conservation is that boundary fluxes vanish.
This holds, for instance, when $M$ is compact without boundary, or when $\rho$ decays rapidly so that
$\int_{\partial U}\rho\,\iota_X\Omega \to 0$ as $U \uparrow M$.

Finally, Eq.~\eqref{eq:continuity_intrinsic} can be written in material form:
\begin{equation}
\left(\pdv{\rho}{t}+X[\rho]\right)=-\rho\,\mathrm{div}_{\Omega}X.
\label{eq:material_form}
\end{equation}
This shows explicitly that $\rho$ is constant along trajectories if and only if the flow is
incompressible with respect to $\Omega$, that is, $\mathrm{div}_{\Omega}X = 0$. In the next section we
prove that Hamiltonian flows satisfy precisely this Liouville incompressibility, leading to Liouville’s
equation $\partial_t\rho+\{\rho,H\}=0$.


\section{Hamiltonian Incompressibility and Liouville's Equation}

We now specialize to Hamiltonian flows for a probability density $\rho$ transported by an arbitrary smooth flow $X$ on phase space.
We now specialize to Hamiltonian dynamics and show that the defining relation
$\iota_{X_H}\omega=\dd H$ forces the flow to preserve the symplectic form and therefore the Liouville
volume form. This yields incompressibility in the Liouville sense and reduces
\eqref{eq:continuity_intrinsic} to Liouville's equation.

\subsection{Hamiltonian flows preserve $\omega$ and $\Omega$}

\begin{proposition}\label{prop:symplectic_invariance}
Let $(\mathcal{M},\omega)$ be a symplectic manifold and let $X_H$ be the Hamiltonian vector field uniquely defined by
\eqref{eq:XH_def}. Then
\begin{equation}
\mathcal{L}_{X_H}\omega=0.
\label{eq:LXHomega0}
\end{equation}
\end{proposition}

\begin{proof}
Cartan's identity gives $\mathcal{L}_{X_H}\omega=\dd(\iota_{X_H}\omega)+\iota_{X_H}(\dd\omega)$.
Using $\iota_{X_H}\omega=\dd H$ and $\dd\omega=0$, we obtain
\[
\mathcal{L}_{X_H}\omega=\dd(\dd H)+\iota_{X_H}(0)=0.
\]
This proves the claim.
\end{proof}

\begin{corollary}\label{cor:liouville_preservation}
Let $\Omega=\omega^N/N!$ be the Liouville volume form. Then the Hamiltonian flow preserves $\Omega$:
\begin{equation}
\mathcal{L}_{X_H}\Omega=0.
\label{eq:LXHOmega0}
\end{equation}
Equivalently,
\begin{equation}
\mathrm{div}_{\Omega}X_H=0.
\label{eq:divXH0}
\end{equation}
\end{corollary}

\begin{proof}
Since $\Omega=\omega^N/N!$, we have
$\mathcal{L}_{X_H}\Omega=\frac{1}{N!}\mathcal{L}_{X_H}(\omega^N)
=\frac{N}{N!}\,(\mathcal{L}_{X_H}\omega)\wedge\omega^{N-1}$.
By Proposition~\ref{prop:symplectic_invariance}, $\mathcal{L}_{X_H}\omega=0$, hence
$\mathcal{L}_{X_H}\Omega=0$. The equivalence with $\mathrm{div}_{\Omega}X_H=0$ follows from the definition
\eqref{eq:divOmega_def}.
\end{proof}

In canonical coordinates, $\Omega=\dd^N q\,\dd^N p$, and $\mathrm{div}_{\Omega}X_H=0$ reduces to the familiar
vanishing Euclidean divergence of the phase-space velocity field. The formulation in terms of Lie derivatives,
however, is coordinate-independent and makes clear that volume preservation is a geometric consequence of
Hamiltonian evolution.

\subsection{From continuity to Liouville's equation}

Specializing \eqref{eq:continuity_intrinsic} to $X=X_H$ and using \eqref{eq:divXH0}, we obtain
\begin{equation}
\pdv{\rho}{t}+\mathrm{div}_{\Omega}(\rho X_H)=0
\quad\Longrightarrow\quad
\pdv{\rho}{t}+X_H[\rho]=0.
\label{eq:liouville_intrinsic}
\end{equation}
To express $X_H[\rho]$ in bracket form, recall that the Poisson bracket is given by \eqref{eq:PB_general}
and that $X_H$ acts on observables as
\begin{equation}
X_H[F]=\{F,H\}.
\label{eq:XH_action}
\end{equation}
(Indeed, this follows immediately from \eqref{eq:XH_def} and the definition of the Poisson tensor.)

Applying \eqref{eq:XH_action} to $F=\rho$ yields Liouville's equation:
\begin{equation}
\pdv{\rho}{t}+\{\rho,H\}=0.\
\label{eq:Liouville_equation}
\end{equation}

\subsection{Canonical form and the origin of the minus sign}

In canonical coordinates with $\omega=\sum_i \dd q^i\wedge \dd p_i$, the defining relation
$\iota_{X_H}\omega=\dd H$ implies
\begin{equation}
\dot q^i=\pdv{H}{p_i},
\qquad
\dot p_i=-\pdv{H}{q^i}.
\label{eq:Hamilton_eqs_sec3}
\end{equation}
The relative minus sign is fixed by the antisymmetry of $\omega$ and the canonical orientation in
$\dd q^i\wedge \dd p_i$, not by an auxiliary convention.

Substituting \eqref{eq:Hamilton_eqs_sec3} into the canonical continuity equation \eqref{eq:continuity_canonical}
reproduces the coordinate version of \eqref{eq:Liouville_equation}:
\begin{equation}
\pdv{\rho}{t}
+\sum_{i=1}^N\left(
\pdv{H}{p_i}\pdv{\rho}{q^i}
-\pdv{H}{q^i}\pdv{\rho}{p_i}
\right)=0.
\label{eq:Liouville_canonical}
\end{equation}

\subsection{Harmonic oscillator}

For the one-dimensional harmonic oscillator,
\begin{equation}
H(q,p)=\frac{p^2}{2m}+\frac{1}{2}m\omega^2 q^2,
\label{eq:HO_H}
\end{equation}
Hamilton's equations give $\dot q=p/m$ and $\dot p=-m\omega^2 q$. Liouville's equation becomes
\begin{equation}
\pdv{\rho}{t}+\frac{p}{m}\pdv{\rho}{q}-m\omega^2 q\,\pdv{\rho}{p}=0,
\label{eq:HO_Liouville}
\end{equation}
showing that $\rho$ is advected along the closed energy contours $H(q,p)=E$ without compression,
in direct correspondence with $\mathcal{L}_{X_H}\Omega=0$.

\subsection{Jacobian form}
Let $\Phi_t$ be the Hamiltonian flow and write locally $\xi(t)=\Phi_t(\xi_0)$. Preservation of $\Omega$ implies
$\Phi_t^\ast\Omega=\Omega$, and in canonical coordinates this is equivalent to
\begin{equation}
\det\!\left(\pdv{\xi(t)}{\xi_0}\right)=1,
\label{eq:Jacobian_one}
\end{equation}
another common expression of Liouville's theorem.


\section{Dissipation, Fine-Grained Information, and the Classical--Quantum Bridge}

Section~3 showed that Hamiltonian evolution preserves the Liouville volume form
$\Omega=\omega^N/N!$ and thus yields an incompressible phase-space advection of the ensemble density
$\rho$. We now clarify how these statements change for non-Hamiltonian (e.g.\ dissipative) flows and
formalize the sense in which ``fine-grained information'' is conserved.

\subsection{Dissipative dynamics: continuity holds, Liouville does not}

Let $X$ be a general smooth phase-space vector field. The statistical evolution of an ensemble is
still governed by the intrinsic continuity equation
\begin{equation}
\pdv{\rho}{t}+\mathrm{div}_{\Omega}(\rho X)=0,
\label{eq:sec4_continuity_general}
\end{equation}
which expresses local conservation of probability and (under suitable boundary conditions) preserves
normalization.

What typically fails in dissipative systems is the invariance of the Liouville measure:
\begin{equation}
\mathcal{L}_{X}\Omega \neq 0
\qquad\Longleftrightarrow\qquad
\mathrm{div}_{\Omega}X \neq 0.
\label{eq:sec4_nonliouville}
\end{equation}
Using the material form \eqref{eq:material_form}, we obtain
\begin{equation}
\left(\pdv{\rho}{t}+X[\rho]\right)=-\rho\,\mathrm{div}_{\Omega}X,
\label{eq:sec4_material_general}
\end{equation}
so the density carried by a moving phase-space ``fluid element'' changes precisely when the flow is
compressible with respect to $\Omega$.

Equation \eqref{eq:sec4_continuity_general} is a kinematic conservation law. Hamiltonian flows are a
distinguished subclass for which $\mathcal{L}_{X_H}\Omega=0$ (Section~3), implying
$\mathrm{div}_{\Omega}X_H=0$ and hence $D\rho/Dt=0$. For dissipative dynamics one often has
$\mathrm{div}_{\Omega}X<0$ (phase-volume contraction), but the continuity equation remains valid and
ensures that total probability is conserved.

\subsection{Fine-grained information and Gibbs--Shannon entropy}

A standard measure of fine-grained information for a classical ensemble is the Gibbs--Shannon entropy
of $\rho$ relative to $\Omega$:
\begin{equation}
S[\rho]:=-\int_\mathcal{M} \rho\ln\rho\;\Omega,
\label{eq:sec4_entropy_def}
\end{equation}
assuming $\rho\ge 0$ and $\rho\ln\rho$ integrable. The continuity equation guarantees normalization,
but entropy conservation requires an additional geometric condition.

\begin{theorem}\label{thm:sec4_entropy}
Let $\rho$ evolve according to the continuity equation \eqref{eq:sec4_continuity_general} for a flow $X$.
Assume either (i) $\mathcal{M}$ is compact without boundary, or (ii) $\rho$ and $X$ satisfy boundary/decay
conditions such that boundary flux terms vanish. If the flow preserves the Liouville volume form,
$\mathcal{L}_{X}\Omega=0$ (equivalently $\mathrm{div}_{\Omega}X=0$), then
\begin{equation}
\frac{\dd S}{\dd t}=0.
\label{eq:sec4_entropy_conservation}
\end{equation}
\end{theorem}

\begin{proof}
Differentiate \eqref{eq:sec4_entropy_def}:
\[
\frac{\dd S}{\dd t}=-\int_\mathcal{M} (\partial_t\rho)(1+\ln\rho)\,\Omega.
\]
If $\mathrm{div}_{\Omega}X=0$, then \eqref{eq:sec4_continuity_general} becomes $\partial_t\rho+X[\rho]=0$.
Hence
\[
\frac{\dd S}{\dd t}=\int_\mathcal{M} X[\rho](1+\ln\rho)\,\Omega=\int_\mathcal{M} X[\rho\ln\rho]\,\Omega.
\]
Under the stated boundary assumptions and $\mathcal{L}_X\Omega=0$, integration by parts (or equivalently,
$\int_\mathcal{M} \mathcal{L}_X(f\Omega)=0$) yields $\int_\mathcal{M} X[f]\Omega=0$ for suitable $f$, implying $\dd S/\dd t=0$.
\end{proof}

\begin{corollary}
For $X=X_H$, Liouville's theorem (Corollary~\ref{cor:liouville_preservation}) implies
$\mathcal{L}_{X_H}\Omega=0$, hence $S[\rho]$ is conserved.
\end{corollary}

The conservation in Theorem~\ref{thm:sec4_entropy} is fine-grained. It refers to the exact density $\rho$
transported by the flow. Effective entropy increase in macroscopic systems arises from coarse-graining,
noise, collisions, or open-system effects, even when the underlying microscopic dynamics is Hamiltonian.

\subsection{A compressible illustration}

As a simple schematic example of a non-Liouville flow, consider in one degree of freedom the damped
equations
\begin{equation}
\dot q=\frac{p}{m},
\qquad
\dot p=-m\omega^2 q-\gamma p,
\label{eq:sec4_damped}
\end{equation}
which correspond to a vector field $X$ on the $(q,p)$ plane. In canonical coordinates,
$\Omega=\dd q\,\dd p$ and therefore
\begin{equation}
\mathrm{div}_{\Omega}X=\pdv{\dot q}{q}+\pdv{\dot p}{p}=0-\gamma=-\gamma<0,
\label{eq:sec4_damped_div}
\end{equation}
so phase-space volumes contract exponentially. Equation \eqref{eq:sec4_material_general} implies that
$\rho$ is not materially conserved and Theorem~\ref{thm:sec4_entropy} does not apply: fine-grained entropy
need not remain constant.

\subsection{Classical--quantum correspondence: Poisson brackets and commutators}

Hamiltonian evolution of observables can be written compactly as
\begin{equation}
\dot F=\{F,H\},
\label{eq:sec4_classical_heis}
\end{equation}
mirroring the Heisenberg equation in quantum mechanics,
\begin{equation}
\dot{\hat F}=\frac{i}{\hbar}[\hat H,\hat F].
\label{eq:sec4_quantum_heis}
\end{equation}
In the semiclassical correspondence,
\begin{equation}
\frac{1}{i\hbar}[\cdot,\cdot]\ \longleftrightarrow\ \{\cdot,\cdot\}.
\label{eq:sec4_correspondence}
\end{equation}

At the statistical level, Liouville's equation \eqref{eq:Liouville_equation} parallels the von Neumann
equation for the density operator $\hat\rho$,
\begin{equation}
i\hbar\,\partial_t\hat\rho=[\hat H,\hat\rho],
\label{eq:sec4_vonneumann}
\end{equation}
emphasizing that the bracket structure governs time evolution in both classical and quantum
statistical mechanics.

For closed quantum systems, unitary evolution preserves $\mathrm{Tr}(\hat\rho)$ and the von Neumann entropy
$S_{\mathrm{vN}}=-\mathrm{Tr}(\hat\rho\ln\hat\rho)$, mirroring the information-preserving character of
Liouville-measure-preserving classical flows.


\section{Symmetries, Conserved Quantities, and the Bargmann Extension}

The symplectic formulation makes symmetry principles particularly transparent: canonical symmetries
are precisely those transformations preserving the symplectic structure, and conserved quantities
are those observables that Poisson-commute with the Hamiltonian. We summarize the main statements
and illustrate them with the nonrelativistic free particle, where the symmetry algebra closes into
the centrally extended Galilei (Bargmann) algebra.

\subsection{Canonical transformations and symplectic vector fields}

A diffeomorphism $\Phi:\mathcal{M}\to \mathcal{M}$ is a \emph{symplectomorphism} (canonical transformation) if
\begin{equation}
\Phi^\ast\omega=\omega.
\label{eq:sec5_symplecto}
\end{equation}
Infinitesimally, a vector field $X$ generates a one-parameter family of symplectomorphisms iff
\begin{equation}
\mathcal{L}_X\omega=0.
\label{eq:sec5_symplecticVF}
\end{equation}
Such vector fields are called \emph{symplectic}. When $X$ is Hamiltonian, $X=X_G$ for some generator
$G$ satisfying $\iota_{X_G}\omega=\dd G$.

On general symplectic manifolds, not every symplectic vector field is globally Hamiltonian; the obstruction
is cohomological. On the standard phase space $\mathcal{M}=\mathbb{R}^{2N}$ with canonical coordinates and suitable
boundary conditions, one may treat symplectic and Hamiltonian generators as equivalent for the purposes of
the present discussion.

\subsection{Conserved quantities in Poisson form}

For an observable $G(\xi,t)$,
\begin{equation}
\dot G=\{G,H\}+\pdv{G}{t}.
\label{eq:sec5_Gdot}
\end{equation}
Therefore:

\begin{proposition}\label{prop:sec5_conservation}
If $G$ has no explicit time dependence, then $G$ is conserved along the Hamiltonian flow iff
\begin{equation}
\{G,H\}=0.
\label{eq:sec5_conservation_condition}
\end{equation}
\end{proposition}

\begin{proof}
With $\partial_t G=0$, Eq.~\eqref{eq:sec5_Gdot} reduces to $\dot G=\{G,H\}$, so $\dot G=0$ iff $\{G,H\}=0$.
\end{proof}

A Hamiltonian generator $G$ produces an infinitesimal canonical transformation via
$\delta_\epsilon F=\epsilon\{F,G\}$. In this sense Poisson brackets encode symmetry actions, paralleling
the role of commutators in quantum theory.

\subsection{A compact template for symmetry algebras}

A family of generators $\{G_A\}$ closes under the Poisson bracket if
\begin{equation}
\{G_A,G_B\}=f_{AB}^{\ \ C}\,G_C + c_{AB},
\label{eq:sec5_closure_template}
\end{equation}
where $f_{AB}^{\ \ C}$ are structure constants and $c_{AB}$ are possible central terms. The Jacobi identity
of the Poisson bracket implies the Jacobi identity for the generator algebra.

\subsection{General solution of Liouville's equation}
\label{subsec:general_Liouville}

Once the dynamics are fixed, the time evolution of any integral of motion
\(Z(\mathbf{q},\mathbf{p},t)\) is governed by Liouville's equation,
\begin{equation}
  \frac{\partial Z}{\partial t} + \{Z,H\} = 0,
  \label{eq:Liouville_Z}
\end{equation}
where \(\{\cdot,\cdot\}\) is the canonical Poisson bracket in 3D,
\begin{equation}
  \{F,G\} = 
  \sum_{i=1}^3 
  \left(
    \frac{\partial F}{\partial q_i}\frac{\partial G}{\partial p_i}
    -
    \frac{\partial F}{\partial p_i}\frac{\partial G}{\partial q_i}
  \right).
  \label{eq:PB_def}
\end{equation}

In this section we restrict ourselves to Hamiltonians depending only on the
momentum,
\begin{equation}
  H = H(\mathbf{p}),
  \label{eq:H_of_p_only}
\end{equation}
which encompasses, in particular, the free particle
\(H = \mathbf{p}^2/2m\) as the simplest nontrivial example.

With \eqref{eq:H_of_p_only}, Liouville's equation
\eqref{eq:Liouville_Z} becomes the linear first–order PDE
\begin{equation}
  \frac{\partial Z}{\partial t}
  + \sum_{i=1}^3 \frac{\partial H}{\partial p_i}\,
                    \frac{\partial Z}{\partial q_i}
  = 0.
  \label{eq:Liouville_H_of_p}
\end{equation}
Using the method of characteristics, we obtain the characteristic system
\begin{equation}
  \dot{q}_i = \frac{\partial H}{\partial p_i}, 
  \qquad
  \dot{p}_i = 0,
  \qquad
  \dot{t} = 1,
  \qquad
  \dot{Z} = 0.
  \label{eq:char_system}
\end{equation}
Hence the momentum is constant along the characteristics,
\(\mathbf{p}(t)=\mathbf{u}=\mathrm{const}\), and the position evolves as
\(\mathbf{q}(t)=\mathbf{k} + \mathbf{v}(\mathbf{u})\,t\), where
\(\mathbf{v}(\mathbf{u}) := \nabla_{\mathbf{p}} H(\mathbf{u})\) is the
group velocity. It is convenient to introduce the two vector invariants
\begin{equation}
  \mathbf{u} := \mathbf{p},
  \qquad
  \mathbf{k} := \mathbf{q} - \mathbf{v}(\mathbf{p})\,t,
  \label{eq:u_k_def}
\end{equation}
which are constant along each Hamiltonian trajectory.

The general solution of \eqref{eq:Liouville_H_of_p} can then be written as
an arbitrary smooth function of these invariants:
\begin{equation}
  Z(\mathbf{q},\mathbf{p},t) = \mathcal{F}(\mathbf{u},\mathbf{k})
  = \mathcal{F}\!\left(
      \mathbf{p},\,
      \mathbf{q}-\mathbf{v}(\mathbf{p})\,t
    \right).
  \label{eq:Liouville_general_solution}
\end{equation}
For the free particle, \(H=\mathbf{p}^2/2m\), we have
\(\mathbf{v}(\mathbf{p})=\mathbf{p}/m\), so that
\begin{equation}
  \mathbf{u} = \mathbf{p},
  \qquad
  \mathbf{k} = \mathbf{q} - \frac{\mathbf{p}}{m}\,t.
  \label{eq:u_k_free}
\end{equation}
We shall use this free–particle realization to make the ensuing
conservation laws explicit.


\subsection{Unification of conservation laws via Taylor expansion}
\label{subsec:unification_conservation}

The representation \eqref{eq:Liouville_general_solution} shows that
\emph{all} integrals of motion for a Hamiltonian \(H(\mathbf{p})\) are
functions of the invariants \(\mathbf{u}\) and \(\mathbf{k}\). To exhibit
the usual conserved quantities and their algebra in a unified way, we
perform a Taylor expansion of \(\mathcal{F}(\mathbf{u},\mathbf{k})\) around
\(\mathbf{u}=\mathbf{0}\), \(\mathbf{k}=\mathbf{0}\):
\begin{equation}
\begin{split}
  \mathcal{F}(\mathbf{u},\mathbf{k})
  &= c_0
     + a_i u_i + b_i k_i \\
  &\quad
     + \frac{1}{2} A_{ij} u_i u_j
     + B_{ij} u_i k_j
     + \frac{1}{2} C_{ij} k_i k_j
     + \cdots,
\end{split}
\label{eq:F_Taylor}
\end{equation}
where \(c_0\) is a constant, \(a_i,b_i\) are constant vectors, and
\(A_{ij},B_{ij},C_{ij}\) are constant matrices (we sum over repeated
indices). Higher–order terms (cubic and beyond) are omitted here but could
be analyzed in the same spirit.

Inserting \eqref{eq:u_k_free} for the free particle,
each term in \eqref{eq:F_Taylor} yields a conserved quantity. The usual
conservation laws emerge as follows.

\paragraph*{(i) Zeroth order.}
The constant term \(c_0\) represents trivial integrals of motion
(addition of constants).

\paragraph*{(ii) Linear terms.}
The linear part
\begin{equation}
  Z^{(1)} = a_i u_i + b_i k_i
          = \mathbf{a}\cdot\mathbf{p}
          + \mathbf{b}\cdot\left(\mathbf{q}-\frac{\mathbf{p}}{m}t\right)
\end{equation}
produces two independent families of conserved vectors,
\begin{align}
\text{from the $u$ term;}\quad & \mathbf{P}:=\mathbf{p} && \text{(linear momentum)},\\
\text{from the $k$ term;}\quad & \mathbf{K}:=m\mathbf{q}-t\,\mathbf{p} && \text{(Galilean boosts)}.
\end{align}

Here we chose \(\mathbf{a}\) and \(\mathbf{b}\) as basis vectors in
\(\mathbb{R}^3\) so that the components \(P_i\) and \(K_i\) are recovered.

\paragraph*{(iii) Quadratic and bilinear terms.}
Quadratic contributions in \eqref{eq:F_Taylor} unify the energy,
angular momentum, and additional scale–type invariants.

\smallskip
\emph{(a) Symmetric term in \(\mathbf{u}\).}
\begin{equation}
  Z^{(2)}_{uu} = \frac{1}{2} A_{ij} u_i u_j.
\end{equation}
For rotationally invariant \(H(\mathbf{p})\), the natural choice is
\(A_{ij} \propto \delta_{ij}\), giving
\[
  Z^{(2)}_{uu} \propto \mathbf{u}^2 = \mathbf{p}^2,
\]
which is proportional to the Hamiltonian itself in the free case:
\[
  H = \frac{\mathbf{p}^2}{2m}.
\]
Thus the energy arises as the rotationally symmetric quadratic invariant in
\(\mathbf{u}\).

\smallskip
\emph{(b) Antisymmetric mixed term in \(\mathbf{u},\mathbf{k}\).}
Decompose the mixed matrix \(B_{ij}\) into symmetric and antisymmetric
parts,
\(B_{ij} = B_{(ij)} + B_{[ij]}\).
The antisymmetric part can be written in terms of a vector
\(\boldsymbol{\gamma}\) via
\(B_{[ij]} = \varepsilon_{ijk}\gamma_k\). The corresponding contribution is
\begin{equation}
  Z^{(2)}_{uk,\text{antisym}} 
  = B_{[ij]} u_i k_j
  = \boldsymbol{\gamma}\cdot(\mathbf{k}\times \mathbf{u}).
\end{equation}
Choosing a basis for \(\boldsymbol{\gamma}\), this yields the angular
momentum vector:
\begin{equation}
  \mathbf{L} 
  := \mathbf{k}\times\mathbf{u}
  = \left(\mathbf{q}-\frac{\mathbf{p}}{m}t\right)\times\mathbf{p}
  = \mathbf{q}\times\mathbf{p},
\end{equation}
since \(\mathbf{p}\times\mathbf{p}=0\). This recovers conservation of
angular momentum.

\smallskip
\emph{(c) Symmetric mixed term in \(\mathbf{u},\mathbf{k}\):}
The symmetric part \(B_{(ij)}\) contains a scalar invariant of the form
\begin{equation}
  Z^{(2)}_{uk,\text{sym}} = B_{(ij)} u_i k_j.
\end{equation}
For \(B_{(ij)}\propto \delta_{ij}\) one obtains
\[
  Z^{(2)}_{uk,\text{sym}} \propto \mathbf{k}\cdot\mathbf{u}
  = \left(\mathbf{q}-\frac{\mathbf{p}}{m}t\right)\cdot\mathbf{p}
  = \mathbf{q}\cdot\mathbf{p} - \frac{\mathbf{p}^2}{m}\,t.
\]
Using \(H=\mathbf{p}^2/2m\), this becomes
\begin{equation}
  D := \mathbf{p}\cdot\mathbf{q} - 2 H t,
\end{equation}
which is the standard \emph{dilation generator}, associated with scale
transformations in the free theory.

\smallskip
\emph{(d) Quadratic term in \(\mathbf{k}\):}
Finally, the symmetric quadratic form in \(\mathbf{k}\),
\begin{equation}
  Z^{(2)}_{kk} = \frac{1}{2} C_{ij} k_i k_j,
\end{equation}
for \(C_{ij}\propto\delta_{ij}\) produces
\[
  Z^{(2)}_{kk} \propto \mathbf{k}^2
  = \left(\mathbf{q}-\frac{\mathbf{p}}{m}t\right)^2.
\]
Up to normalization, we may define
\begin{equation}
  K_{\mathrm{conf}} := \mathbf{k}\cdot\mathbf{k}
  = \left(\mathbf{q}-\frac{\mathbf{p}}{m}t\right)^2,
\end{equation}
which plays the role of a \emph{special conformal generator} associated
with expansions in the free theory  \cite{GomisEtAlPRL2020}.

Higher–order terms in the Taylor series \eqref{eq:F_Taylor} generate
higher–rank conserved tensors. For the purposes of this work, it suffices
to focus on the linear and quadratic sectors, which already contain the
standard conserved quantities of nonrelativistic mechanics and the basic
scale–type generators.


\subsection{Extended algebra and central charge}
\label{subsec:extended_algebra}

We now summarize the Poisson–bracket algebra satisfied by the conserved
generators obtained above. For the free Hamiltonian
\(H=\mathbf{p}^2/2m\), the fundamental generators in phase space are:
\begin{align}
  P_i &= p_i,
  &&\text{(linear momentum)}, \\
  J_i &= \varepsilon_{ijk} q_j p_k,
  &&\text{(angular momentum)}, \\
  K_i &= m q_i - t p_i,
  &&\text{(Galilean boosts)}, \\
  H &= \frac{\mathbf{p}^2}{2m},
  &&\text{(energy)}, \\
  D &= \mathbf{p}\cdot\mathbf{q} - 2Ht,
  &&\text{(dilation)}, \\
  K_{\mathrm{conf}} &= \left(\mathbf{q}-\frac{\mathbf{p}}{m}t\right)^2,
  &&\text{(special conformal)}, \\
  M &= m,
  &&\text{(mass, central charge)}.
\end{align}
The mass $M$ Poisson–commutes with all generators,
\(\{M,\cdot\}=0\).

Using the canonical bracket \eqref{eq:PB_def}, one finds the nonvanishing Poisson brackets:

{\small
\begin{subequations}
\label{eq:extended_algebra}
\begin{alignat}{3}
\{J_i,J_j\} &= \varepsilon_{ijk} J_k,
&
\{J_i,P_j\} &= \varepsilon_{ijk} P_k,
&
\{J_i,K_j\} &= \varepsilon_{ijk} K_k,
  \\[0.25em]
  \{H,K_i\} &= -P_i,
  &
  \{K_i,P_j\} &= M\,\delta_{ij},
  &
  \{P_i,P_j\} &= 0,
  \\[0.25em]
  \{D,H\} &= 2H,
  &
  \{D,P_i\} &= P_i,
  &
  \{D,K_i\} &= -K_i,
  \\[0.25em]
  \{K_{\mathrm{conf}},K_i\} &= 0.
\end{alignat}
\end{subequations}
}

In addition, the brackets involving the special conformal generator have the expected
Schr\"odinger-type structure. Up to an overall normalization, fixed below, $K_{\mathrm{conf}}$
brackets with $H$ into the dilatation generator and maps translations into boosts. Equivalently,
$\{H,K_{\mathrm{conf}}\}\sim D$ and $\{K_{\mathrm{conf}},P_i\}\sim K_i$.

All brackets with $M$ vanish and the remaining ones are zero. If we treat the mass as a central
element, we obtain the Bargmann algebra, which is a central extension of the Galilean algebra. The
appearance of $m$ as a central term explains why mass behaves as a superselected parameter in
nonrelativistic quantum theory, and it provides a clean classical prelude to the corresponding quantum
commutator algebra. Related modern developments on symmetry structures arising from post-Galilean
expansions and other nonrelativistic limits can be found in Ref.~\cite{GomisEtAlPRL2020}.

We now choose a convenient normalization for the special conformal
generator so that the triplet $(H,D,K_{\text{conf}})$ closes exactly the
standard $\mathfrak{sl}(2,\mathbb{R})$ subalgebra of the Schr\"odinger
group. Instead of $(\mathbf{q}-\mathbf{p}t/m)^2$, we define
\begin{equation}
  K_{\mathrm{conf}} := C
  = -\,\frac{m}{2}\left(\mathbf{q}-\frac{\mathbf{p}}{m}t\right)^2,
  \label{eq:Kconf_normalized}
\end{equation}
while keeping
\begin{equation}
  H = \frac{\mathbf{p}^2}{2m},
  \qquad
  D = \mathbf{p}\cdot\mathbf{q} - 2Ht.
  \label{eq:HD_defs}
\end{equation}
A direct computation with the canonical Poisson bracket
\eqref{eq:PB_def} yields
\begin{equation}
  \{D,H\} = 2H,
  \
  \{D,K_{\mathrm{conf}}\} = -2K_{\mathrm{conf}},
  \
  \{H,K_{\mathrm{conf}}\} = D,
  \label{eq:sl2R_schrodinger}
\end{equation}
so that $(H,D,K_{\mathrm{conf}})$ form a classical $\mathfrak{sl}(2,\mathbb{R})$
triple. This is the familiar $SO(2,1)$ (or $SL(2,\mathbb{R})$) sector of the
Schr\"odinger symmetry, acting on time as time translations $(H)$,
dilatations $(D)$, and special conformal transformations
$(K_{\mathrm{conf}})$.

The key point for our purposes is that the Poisson--bracket algebra
\eqref{eq:extended_algebra} emerges \emph{directly} from the general
solution of Liouville's equation \eqref{eq:Liouville_general_solution}
once we organize the invariants in a Taylor expansion
\eqref{eq:F_Taylor}. In this way, the conservation of linear momentum,
boosts, angular momentum, energy, and the scale--type generators $D$ and
$K_{\mathrm{conf}}$ are all unified as different polynomial sectors of a
single function $\mathcal{F}(\mathbf{u},\mathbf{k})$.



\section{Extended Phase Space and Hamilton--Jacobi as an Invariance Condition}

The Hamilton--Jacobi (HJ) equation is often introduced as an alternative formulation of Hamiltonian
dynamics based on generating functions. In the symplectic framework, it can be presented cleanly as
a compatibility (invariance) condition on a Lagrangian submanifold. The extended phase-space language
makes the logic particularly transparent.

\subsection{Extended phase space and the canonical one-form}

Consider the \emph{extended phase space}
\begin{equation}
\mathcal{M}_{\mathrm{ext}} := \mathcal{M} \times \mathbb{R}_t \times \mathbb{R}_{p_t},
\qquad
(\xi^a,t,p_t)\in \mathcal{M}_{\mathrm{ext}},
\label{eq:ext_phase_space}
\end{equation}
where $(t,p_t)$ is treated as an additional canonical pair. Define the extended canonical one-form
\begin{equation}
\Theta_{\mathrm{ext}} := \sum_{i=1}^N p_i\,\dd q^i + p_t\,\dd t,
\label{eq:Theta_ext}
\end{equation}
and the associated extended symplectic form
\begin{equation}
\omega_{\mathrm{ext}} := -\dd \Theta_{\mathrm{ext}}
= \sum_{i=1}^N \dd q^i\wedge \dd p_i + \dd t\wedge \dd p_t.
\label{eq:omega_ext}
\end{equation}
(Here we adopt the standard convention $\omega=-\dd\Theta$; the sign convention is immaterial as long as
it is used consistently.)

If needed, the extended Liouville volume form is
\begin{equation}
\Omega_{\mathrm{ext}}:=\frac{\omega_{\mathrm{ext}}^{\,N+1}}{(N+1)!},
\label{eq:Omega_ext}
\end{equation}
but for the HJ argument below, $\omega_{\mathrm{ext}}$ is the key object.

\subsection{Extended Hamiltonian and Hamiltonian flow}

Let the physical Hamiltonian depend on $(q,p,t)$, i.e.\ $H=H(q,p,t)$. Define the extended Hamiltonian
constraint function
\begin{equation}
\mathcal{H}(q,p,t,p_t):=H(q,p,t)+p_t.
\label{eq:calH}
\end{equation}
The Hamiltonian vector field $X_{\mathcal{H}}$ on $(M_{\mathrm{ext}},\omega_{\mathrm{ext}})$ is defined by
\begin{equation}
\iota_{X_{\mathcal{H}}}\omega_{\mathrm{ext}}=\dd \mathcal{H}.
\label{eq:XcalH_def}
\end{equation}

In canonical coordinates, the corresponding equations are
\begin{equation}
\label{eq:ext_canonical_eqs}
\begin{aligned}
\dot q^i &= \pdv{\mathcal{H}}{p_i} = \pdv{H}{p_i},
\qquad
\dot p_i &= -\,\pdv{\mathcal{H}}{q^i} = -\,\pdv{H}{q^i},
\\[0.3em]
\dot t   &= \pdv{\mathcal{H}}{p_t} = 1,
\qquad
\dot p_t &= -\,\pdv{\mathcal{H}}{t} = -\,\pdv{H}{t}.
\end{aligned}
\end{equation}

Thus the extended flow reproduces the original Hamilton equations while treating time as a canonical variable.

\subsection{Hamilton--Jacobi from a Lagrangian submanifold}

Let $S(q,t)$ be a generating function and consider the submanifold $\Lambda_S\subset \mathcal{M}_{\mathrm{ext}}$
defined by
\begin{equation}
p_i=\pdv{S}{q^i},
\qquad
p_t=\pdv{S}{t}.
\label{eq:LambdaS_def}
\end{equation}
The pullback of $\Theta_{\mathrm{ext}}$ to $\Lambda_S$ is
\[
\Theta_{\mathrm{ext}}\big|_{\Lambda_S}
=\sum_i \pdv{S}{q^i}\dd q^i + \pdv{S}{t}\dd t
=\dd S,
\]
hence $\omega_{\mathrm{ext}}\big|_{\Lambda_S}=-\dd(\Theta_{\mathrm{ext}}|_{\Lambda_S})=-\dd(\dd S)=0$.
Therefore $\Lambda_S$ is a \emph{Lagrangian} submanifold of $(\mathcal{M}_{\mathrm{ext}},\omega_{\mathrm{ext}})$.

The Hamilton--Jacobi equation is obtained by imposing that $\Lambda_S$ lies on the constraint surface
$\mathcal{H}=0$:
\begin{equation}
\mathcal{H}\big(q,\partial_q S,t,\partial_t S\big)=0
\quad\Longleftrightarrow\quad
\pdv{S}{t}+H\!\left(q,\pdv{S}{q},t\right)=0.\
\label{eq:HJ_equation}
\end{equation}

Geometrically, the HJ equation expresses the compatibility of the Lagrangian graph $\Lambda_S$ with the
Hamiltonian evolution in extended phase space. One may equivalently view it as the condition that the
Hamiltonian flow generated by $\mathcal{H}$ is tangent to the submanifold defined by \eqref{eq:LambdaS_def},
so that the dynamics can be reduced to the evolution of the generating function $S(q,t)$.


\section{Summary and Discussion}
\label{sec:discussion}

The main pedagogical goal of this work is to reorganize a standard topic---Liouville's theorem and
Hamiltonian dynamics---around two structural elements that are already implicit in advanced
treatments of mechanics: (i) the canonical Poisson-bracket relations between position and momentum,
which encode the kinematics of phase space \cite{goldstein,landau}, and (ii) local conservation of
probabilistic information for ensembles \cite{HenrikssonLJP2022}. The continuity equation is not
merely a computational tool; formulated intrinsically with respect to the Liouville volume form
$\Omega=\omega^N/N!$, it expresses probability conservation in a coordinate-independent way and
identifies the appropriate notion of divergence on phase space via
$\mathcal{L}_X\Omega=(\mathrm{div}_{\Omega}X)\Omega$. From the instructor's point of view, this
ordering separates kinematics (canonical Poisson brackets) from statistics (probability transport),
prevents the common conflation between ``continuity'' and ``measure preservation,'' and yields a
short chain of implications that can be assigned as guided derivations in an upper-division
mechanics course.

Within this framework, Hamiltonian dynamics is singled out by geometry \cite{arnold}. Once a
symplectic structure $\omega$ and a Hamiltonian function $H$ are given, the Hamiltonian vector field
$X_H$ is fixed by $\iota_{X_H}\omega=\dd H$. Cartan's identity then yields $\mathcal{L}_{X_H}\omega=0$
and therefore $\mathcal{L}_{X_H}\Omega=0$, so ensemble densities are transported by an
$\Omega$-incompressible phase-space flow \cite{pathria}. In canonical coordinates this reduction
recovers Liouville's equation and Hamilton's equations, with the sign structure in the momentum
equation traced to the antisymmetry and canonical orientation encoded in
$\omega=\sum_i \dd q^i\wedge \dd p_i$.

This viewpoint also separates two statements that are often conflated in instruction. The continuity
equation secures conservation of total probability under suitable boundary conditions for any smooth
flow, including dissipative ones, whereas preservation of the Liouville measure is a stronger
property that characterizes Hamiltonian evolution. The same separation sharpens the meaning of
``information conservation.'' Fine-grained Gibbs--Shannon entropy is conserved for
Liouville-measure-preserving dynamics \cite{shannon}, while dissipative systems may still obey a
continuity equation even though they fail to preserve $\Omega$.

The symmetry analysis fits naturally into the same bracket-based language. Conserved generators are
characterized by Poisson-commutation with $H$, and their closure under $\{\cdot,\cdot\}$ organizes
canonical symmetries into Lie algebras. For the nonrelativistic free particle, the centrally extended
Galilei algebra (Bargmann) exhibits the mass as a central charge, providing a classical prelude to
projective quantum representations, and modern developments in post-Galilean expansions and related
nonrelativistic limits can be found in Ref.~\cite{GomisEtAlPRL2020}. Finally, the extended phase-space
viewpoint offers a compact route to Hamilton--Jacobi theory by treating time as a canonical
coordinate and interpreting the HJ condition as the compatibility of a Lagrangian submanifold with
the Hamiltonian flow. clarifies the role of time as a canonical
coordinate.

\medskip
The emphasis on Poisson brackets also clarifies the classical--quantum structural bridge
\cite{dirac,sakurai}. Under canonical quantization, the classical Poisson algebra is replaced by a
noncommutative operator algebra, and the operator-based viewpoint on brackets can be formulated in a
useful parallel language \cite{KoidePRA2021}. At the level of dynamics, commutator-based evolution laws
(Heisenberg--von Neumann) mirror Poisson-bracket evolution laws in structure. As a complementary link,
Ehrenfest's theorem explains why expectation values often exhibit classical-looking evolution under
standard Hamiltonians, while genuinely quantum features (noncommutativity and dispersion) persist.

The reconstruction presented here assumes a symplectic phase space and therefore addresses systems
whose dynamics is Hamiltonian (or can be embedded into a Hamiltonian system). While the continuity
equation itself is general, the reduction to Liouville's equation and the conservation of
fine-grained entropy rely on Liouville-measure preservation. Open systems, stochastic forcing, coarse-graining, and phenomenological dissipation require additional
structures, for example contact geometry, metriplectic formalisms, or explicit coupling to reservoirs,
and are beyond the scope of this paper. For recent developments connecting Liouville--Arnold--type results with homogeneous
symplectic and contact Hamiltonian systems, see Ref.~\cite{ColomboEtAlGM2025}. Nevertheless, the
present formulation provides a clean baseline against which such extensions can be contrasted.

\subsection*{Pedagogical outlook}
From an instructional standpoint, the approach suggests a natural sequence of learning outcomes:
(i) interpret the canonical Poisson-bracket relations as the defining kinematic input of Hamiltonian
mechanics, (ii) identify the Liouville measure as the canonical phase-space volume induced by
$\omega$, (iii) understand Liouville's theorem as measure preservation under Hamiltonian evolution,
(iv) read Liouville's equation as probability advection by an incompressible flow, and (v) connect
symmetries to conserved generators via the Poisson algebra. These points can be supported by short guided problems, for instance verifying incompressibility for standard Hamiltonians
or computing the Bargmann brackets explicitly, and by numerical visualizations of advection in phase space.



\begin{thebibliography}{99}

\bibitem{goldstein} 
H. Goldstein, C. Poole, and J. Safko, 
\textit{Classical Mechanics}, 3rd ed. (Addison-Wesley, Reading, MA, 2002).

\bibitem{landau} 
L. D. Landau and E. M. Lifshitz, 
\textit{Mechanics}, Vol. 1 (Butterworth-Heinemann, Oxford, 1976).

\bibitem{shannon} 
C. E. Shannon, 
``A Mathematical Theory of Communication'', 
\textit{Bell System Technical Journal} \textbf{27}, 379--423 (1948).

\bibitem{arnold} 
V. I. Arnold, 
\textit{Mathematical Methods of Classical Mechanics}, 2nd ed. (Springer-Verlag, New York, 1989).

\bibitem{sakurai} 
J. J. Sakurai, 
\textit{Modern Quantum Mechanics}, 2nd ed. (Addison-Wesley, Reading, MA, 1994).

\bibitem{pathria} 
R. K. Pathria and P. D. Beale, 
\textit{Statistical Mechanics}, 3rd ed. (Elsevier, Amsterdam, 2011).

\bibitem{dirac}
P. A. M. Dirac,
\textit{The Principles of Quantum Mechanics}, 4th ed. (Oxford University Press, Oxford, 1958).


\bibitem{GomisEtAlPRL2020}
J. Gomis, A. Kleinschmidt, J. Palmkvist, and P. Salgado-Rebolledo,
``Symmetries of Post-Galilean Expansions,''
\textit{Phys. Rev. Lett.} \textbf{124}, 081602 (2020).
doi:10.1103/PhysRevLett.124.081602


\bibitem{KoidePRA2021}
T. Koide,
``Poisson bracket operator,''
\textit{Phys. Rev. A} \textbf{104}, 042213 (2021).

\bibitem{ColomboEtAlGM2025}
L. Colombo, M. de Le\'{o}n, M. Lainz, and A. L\'{o}pez-Gord\'{o}n,
``Liouville--Arnold theorem for homogeneous symplectic and contact Hamiltonian systems,''
\textit{Geometric Mechanics} \textbf{2}(3), 275--307 (2025).

\bibitem{HenrikssonLJP2022}
A. Henriksson,
``Liouville's theorem and the foundation of classical mechanics,''
\textit{Lithuanian Journal of Physics} \textbf{62}(2), 73--80 (2022).


\end{thebibliography}
\end{document}